\documentclass[12pt]{article}
\usepackage{pdproc,epsfig} 

\begin{document}

\renewcommand{\thefootnote}{\alph{footnote}}

\title{
QUARK-LEPTON COMPLEMENTARITY; A REVIEW~\protect\footnote{
Written version of a talk presented at the XI International Workshop on 
``Neutrino Telescopes'', Venice, Italy, February 22-25, 2005.}} 

\author{HISAKAZU MINAKATA}

\address{Department of Physics, Tokyo Metropolitan University, \\
1-1 Minami-Osawa, Hachioji, Tokyo 192-0397, Japan\\
 {\rm E-mail: minakata@phys.metro-u.ac.jp}}




\abstract{It has been recognized recently that there is a remarkable 
empirical relation between lepton and quark mixing angles, 
$\theta_{12} + \theta_C \approx \pi/4$. If not accidental, 
it should testify for yet uncovered new relationship between 
the fundamental twin particles in nature which only differ in 
their ability to feel color.  
The nontrivial structure which is presumed to exist behind the 
empirical relation is named as ``quark-lepton complementarity''. 
In this talk, I review the idea at the kind request of the organizer.
Starting from pedagogical discussions of bimaximal mixing, 
which likely to be involved in the whole picture, I try to give a flavor 
of the new field which is still in rapid development. 
Toward the more balanced knowledges of flavor mixing in lepton 
and quark sectors, I describe a promising way for precision 
measurement of $\theta_{12}$ which utilizes solar and reactor neutrinos. 
}

\normalsize\baselineskip=15pt

\section{Introduction}

In the last year, three experiments observing neutrinos originated from 
the atmosphere~\cite{SKatm_new}, the reactor~\cite{KamLAND_new}, 
and the accelerator~\cite{K2K_new} all saw the oscillatory behavior, 
providing us with a long awaited confirmation of $\nu$ mass-induced 
neutrino oscillation since its discovery by Super-Kamiokande \cite{SKatm}. 
Now, we can talk about neutrino masses and lepton flavor mixing \cite{MNS} 
with confidence, and it made the {\it by now traditional} workshop series 
``Neutrino Telescopes'' in Venice even more important to establish 
future direction of research in fundamental particle physics. 
I should note that we all owe much to Milla for her tireless great 
enthusiasm for having the meeting in such a scenic place.

Let me start by giving a few ward on the thus uncovered structure 
of lepton flavor mixing; 
It consists of 
a large and possibly maximal angle $\theta_{23}$ (atmospheric angle \cite{SKatm}), 
a large but non-maximal angle $\theta_{12}$ (solar angle \cite{solar}), and 
a known-to-be small angle $\theta_{13}$ (reactor angle \cite{CHOOZ}). 
The rich variety in lepton mixing angles from small to nearly maximal 
mixing is in sharp contrast to quark mixing angles and it must be 
testifying something important on how nature organized the 
structure of, to date, the most fundamental matter. 
One of the key wards in understanding the structure may be the 
notion of lepton-quark correspondence which dates back to 
late fifties and early sixties \cite{QLC_old}. 
The contemporary theory of the fundamental matter, of course, lends 
supports to the relation in the form of anomaly cancellation 
mechanism in the standard model. Therefore, there are enough 
circumstantial evidences of the fact that the existence of leptons 
and quarks is mutually dependent with each others. 
The real problem is, however, to uncover the whole picture of 
how they are related. 
The traditional answer to this problem is, of course, grand unification 
in which quarks are leptons are unified into the same multiplet \cite{GUT}.

Quark-lepton complementarity (QLC) \cite{mina-smi,raidal} is one of the 
approaches along the line of thought \cite{mohapatra}. 
We start from an empirical observation that the solar angle 
$\theta_{\odot} \equiv \theta_{12}$ and the Cabibbo angle $\theta_{C}$ add up 
to $\pi/4$ in a good approximation \cite{NOVE03_smi}; 
\begin{eqnarray}
\theta_{\odot} + \theta_C = 45.1^{\circ} \pm 2.4^{\circ}~~~~~  (1\sigma) .
\label{solar-cabibbo}
\end{eqnarray} 
It appears that it is so close to 
\begin{eqnarray}
\theta_{\odot} + \theta_C = \pi/4, 
\label{QLC}
\end{eqnarray} 
the charming relation which 
suggests that quarks and leptons are hiding their common roots. 
Many questions immediately arise; 
What is the interpretation of the empirical relation? 
Are there particle physics models that naturally embody this relation? 
It is the purpose of this talk to review the status of the new approach, 
a duty which was kindly assigned to me by the organizer. 
My presentation here is meant to be very pedagogical and mainly 
for experimentalists, or people who are trying to get to 
the relatively new idea.
For further references of quark-lepton complementarity (whose 
naming is due to \cite{mina-smi}) may be found in 
\cite{pakvasa,mohapatra1,ramond,QLCmore,mohapatra2,LSS}.

One of the directions which will be pursued in the new era of 
neutrino physics will be precision determination of the lepton 
mixing parameters. 
The approach of QLC and the trend to precision measurement 
are ``complementary'' with each other.
I mean, there is a real need for precision determination of 
$\theta_{12}$ to verify the relationship $\theta_{\odot} + \theta_C = \pi/4$, 
or find the deviation from it. 
Now, I would like to note that the Cabibbo angle is measured 
with great precision of about 1.4\% in 
$\sin^2\theta_{C}$ at 90\% CL \cite{PDG}. 
What about the solar angle? 
It is about 14\% in $\sin^2\theta_{12}$ at the same CL \cite{global}. 
What a large disparity between accuracies in measurement of 
lepton and quark mixing angles!
I am sure that nature feels sad about our {\it unequal treatment} of the twin 
particles she created which differ only by possessing or lacking ability 
of tasting color. 
Therefore, I will try to also cover the question of how and to what 
extent accuracy of determination of $\theta_{12}$ can be improved.

\section{QLC; questions}

First, let me list some immediate questions about QLC.
In fact, there are bunch of them:

\begin{itemize}

\item

Suppose that the QLC relation (\ref{QLC}) is correct. 
Then, the question is; Is there a similar relation 
\begin{eqnarray}
\theta_{23}^{lepton} + \theta_{23}^{quark} = \pi/4 ?
\label{QLC2}
\end{eqnarray} 
The relation is perfectly allowed by the current data;  
$37^{\circ} \leq \theta_{23}^{lepton} \leq 53^{\circ}$~\cite{SKatm}, and 
$2.3^{\circ} \leq \theta_{23}^{quark} \leq 2.5^{\circ}$~\cite{PDG}, 
both at 90\% CL.

\item

What is the reason why the analogous relation 
\begin{eqnarray}
\theta_{13}^{lepton} + \theta_{13}^{quark} = \pi/4 
\label{QLC3}
\end{eqnarray} 
does {\it not} hold? 
Experimentally the sum in (\ref{QLC3}) is less than 8.5$^{\circ}$ 
\cite{global}, far from 45$^{\circ}$.

\item
Suppose that the relations (\ref{QLC}) and (\ref{QLC2}) are 
approximately correct. 
Then, the question is; Are there relationship between the deviation 
from the maximal in (\ref{QLC}) and (\ref{QLC2})?

\item
Is the possible deviation from the maximal of $\theta_{23}$
connected with ``deviation from zero'' of $\theta_{13}$?
(For possible symmetries which lead to such connection, see 
\cite{mu-tau}.)
If so, then, what is the role played by 
a possible deviation from (\ref{QLC}) in the game?

\item
Suppose that there exist well defined models which realize the 
QLC relation (\ref{QLC}). 
Then, the question is; how the relation made stable against the 
changes of parameters of the model?

\end{itemize}

\noindent
Unfortunately, no definitive answer is offered to any of these 
questions at this moment. 
QLC is a brand-new approach and it is in a too premature 
stage to answer them. 
Nonetheless, let me try to give some hints toward motivating 
the real understanding.

\section{What Does It Mean?}

The QLC relation (\ref{QLC}) implies the presence of maximal 
mixing angle in somewhere in (1-2) sector of quark-lepton 
mixing matrix, the MNS matrix \cite{MNS}. 
Given another mixing angle close to the maximal in (2-3) sector, 
$\theta_{atm} \equiv \theta_{23}$, it naturally lead us to the new 
bimaximal mixing hypothesis. 
The old version of bimaximal mixing \cite{bimax} was bravely 
invented by people who coined to the possibility that the 
solar neutrino problem is solved by the vacuum oscillation solution. 
In this viewpoint, the bimaximal mixing would have been the issue 
purely inside the realm of lepton flavor mixing, having nothing to 
do with quark mixing. 

On the contrary, our new bimaximal mixing ansats requires to 
embed at least one of the maximal angles into the ``unified'' 
quark-lepton sectors.
To me, it is one of the most interesting features of QLC ansats; 
QLC requires quark-lepton unification.
Therefore, most probably QLC implies the existence of maximal 
angle which is neither in lepton nor in quark sectors.

\subsection{Pedagogical bimaximal mixing}

By the  bimaximal mixing I mean flavor mixing matrix 
\begin{eqnarray}
U_{bimax} \equiv 
\left[
\begin{array}{ccc}
\frac{1}{\sqrt{2}} & -\frac{1}{\sqrt{2}} & 0 \\
\frac{1}{2} & \frac{1}{2}  & -\frac{1}{\sqrt{2}} \\
\frac{1}{2} & \frac{1}{2} & \frac{1}{\sqrt{2}}
\end{array}
\right]
\end{eqnarray}
From which mass matrix does the bimaximal mixing come?
Assuming that there are no other entities which come into play, 
it is easy to answer the question: 
\begin{eqnarray}
U_{bimax} \left[
\begin{array}{ccc}
m_1 & 0 & 0 \\
0 & m_2 & 0 \\
0 & 0 & m_3
\end{array}
\right]
U_{bimax}^{\dagger} =
\left[
\begin{array}{ccc}
\frac{m_1}{2} + \frac{m_2}{2} & 
\frac{m_1}{2\sqrt{2}} - \frac{m_2}{2\sqrt{2}}  &  
\frac{m_1}{2\sqrt{2}} - \frac{m_2}{2\sqrt{2}}  \\
 \frac{m_1}{2\sqrt{2}} - \frac{m_2}{2\sqrt{2}}  & 
 \frac{m_1}{4} + \frac{m_2}{4} + \frac{m_3}{2} &  
 \frac{m_1}{4} + \frac{m_2}{4} - \frac{m_3}{2}  \\
 \frac{m_1}{2\sqrt{2}} - \frac{m_2}{2\sqrt{2}}  &  
 \frac{m_1}{4} + \frac{m_2}{4} - \frac{m_3}{2}  &  
 \frac{m_1}{4} + \frac{m_2}{4} + \frac{m_3}{2} 
\end{array}
\right]
\end{eqnarray}
You may complain that it is not very illuminating. Yes, you are quite right. 
So, let us examine a bit of simplified cases. 
Suppose that the lightest neutrino mass is much smaller than 
$\sqrt{\Delta m^2_{atm}}$, the hierarchical mass pattern. 
Then, there are three cases, 
one ``normal'' ($m_3 \gg m_2 \approx m_1$)
and two ``inverted'' ($m_2 \approx m_1  \gg m_3$) 
mass hierarchies:
\begin{eqnarray}
\mbox{Normal:}
\hspace{0.5cm}
M_{atm} &\equiv &
U_{bimax}
\left[
\begin{array}{ccc}
0 & 0 & 0 \\
0 & 0 & 0 \\
0 & 0 & m_3
\end{array}
\right]
U_{bimax}^{\dagger} =
\left[
\begin{array}{ccc}
0 & 0 & 0 \\
0 & \frac{m_3}{2} & -\frac{m_3}{2} \\
0 & -\frac{m_3}{2} & \frac{m_3}{2}
\end{array}
\right], \\
\mbox{Inverted I:}
\hspace{0.5cm}
M_{atm} &\equiv &
U_{bimax}
\left[
\begin{array}{ccc}
m_2 & 0 & 0 \\
0 & m_2 & 0 \\
0 & 0 & 0
\end{array}
\right] 
U_{bimax}^{\dagger} =
\left[
\begin{array}{ccc}
m_2 & 0 & 0 \\
0 & \frac{m_2}{2} & \frac{m_2}{2} \\
0 & \frac{m_2}{2} & \frac{m_2}{2}
\end{array}
\right], \\
\mbox{Inverted II:}
\hspace{0.5cm}
M_{atm} &\equiv &
U_{bimax}
\left[
\begin{array}{ccc}
m_2 & 0 & 0 \\
0 & -m_2 & 0 \\
0 & 0 & 0
\end{array}
\right] 
U_{bimax}^{\dagger} =
\left[
\begin{array}{ccc}
0 & \frac{m_2}{\sqrt{2}}  &  \frac{m_2}{\sqrt{2}}  \\
\frac{m_2}{\sqrt{2}} & 0 & 0  \\
\frac{m_2}{\sqrt{2}} & 0 & 0 
\end{array}
\right],
\end{eqnarray}
where 
$m_3 = \sqrt{\Delta m^2_{atm}}$ and 
$m_2 =  \sqrt{\Delta m^2_{atm}}$ in 
the normal and the inverted mass hierarchies, respectively. 
We note that in the above all cases the mass matrices have 
$\mu \leftrightarrow \tau$ exchange symmetry \cite{mu-tau}. 
The Inverted II case has an extra $L_{e} - L_{\mu} - L_{\tau}$ 
symmetry widely discussed in the literature \cite{e-mu-tau}.

\subsection{Perpurbative approach}

One can phenomenologically describe QLC in a perpurbative way 
starting from the mass matrix above, corresponding to each mass 
pattern, as done by Ferrandis and Pakvasa \cite{pakvasa}. 
In the case of Inverted II, 
which is favored by the authors, it takes the form 
$M = M_{atm} + M_{sol} + M_{QLC}$, 
where $M_{solar}$ and $M_{QLC}$ denote the solar scale and the 
QLC corrections, respectively. They are 
\begin{eqnarray}
M_{sol}   = \gamma m_2
\left[
\begin{array}{ccc}
 \frac{1}{2} & -\frac{1}{2\sqrt{2}}  &  -\frac{1}{2\sqrt{2}}  \\
 -\frac{1}{2\sqrt{2}}   & \frac{1}{4} & \frac{1}{4} \\
 -\frac{1}{2\sqrt{2}}  & \frac{1}{4}  & \frac{1}{4} 
\end{array}
\right], 
\hspace{0.3 cm}
M_{QLC} =  \lambda m_2 
\left[
\begin{array}{ccc}
-1  & 0  & 0 \\
0 & \frac{1}{2} & \frac{1}{2} \\
0 &  \frac{1}{2}  & \frac{1}{2}
\end{array}
\right], 
\end{eqnarray}
where 
$\gamma \approx (\Delta m_{\odot}^{2} / \Delta m_{atm}^{2})/2 
\approx 0.016$ and $\lambda \approx \sin\theta_{C}$. 
There are some interesting differences between the three cases. 
The ordering between $M_{solar}$ and $M_{QLC}$ differ in these 
three cases; 
$M_{QLC} < M_{solar}$ in the Inverted I and 
$M_{QLC} \gg M_{solar}$ in the Inverted II cases \cite{pakvasa}.

\section{QLC as Indication of Quark-Lepton Unification}

As I emphasized earlier the most charming features of QLC 
is that it strongly suggests quark-lepton unification in some forms. 
Let me discuss this point in more detail. 
While concrete models which correctly predicted the QLC 
relation (\ref{solar-cabibbo}) prior experimental observation are missing, 
the structure of embedding of the relation into 
GUT-like scenarios, once explicitly formulated, allows us to test 
the QLC embedded GUT-like scenarios experimentally. 
This is the topics thoroughly discussed in \cite{mina-smi}.

Let us sketch the basic points of the discussions in \cite{mina-smi}. 
There are two types of scenarios, depending upon from which sector 
the maximal 1-2 angle comes, 
the lepton-origin bimaximal and the neutrino-origin bimaximal scenarios. 
Let us first recall, not to be confused, the definition of the MNS and 
the CKM matrix. They are 
\begin{eqnarray}
U_{MNS} = U_{lepton}^{\dagger} U_{\nu}, 
\hspace{1cm} 
V_{CKM} = V_{up}^{\dagger}  V_{down}, 
\end{eqnarray}
where $U_{\nu}$ and $U_{lepton}$ denote the matrices which diagonalize 
neutrino and charged lepton mass matrices, respectively 
(and the same as in quarks). 
Then, the ideas behind the both scenarios can be displayed in a simple 
illustrative way as below. 

\begin{itemize}

\item

Lepton-origin bimaximal scenario

\end{itemize}
\begin{eqnarray}
U_{\nu} = V_{CKM}^{\dagger} \hspace{0.5cm} 
& \leftarrow \mbox{GUT} \rightarrow & 
\hspace{0.5cm} 
 V_{up}= V_{CKM}^{\dagger} \\
U_{lepton} = U_{bimaximal} \hspace{0.5cm} 
& \leftarrow \mbox{Lopsided}  \hspace{0.5cm}& 
\hspace{0.5cm}  V_{down} = I 
\end{eqnarray}
where ``Lopsided'' indicates that the lopsided scenario 
\cite{lopsided} may gives the relation pointed by an arrow. 
\begin{itemize}
\item
Neutrino-origin bimaximal scenario
\end{itemize}
\begin{eqnarray}
U_{\nu} = U_{bimaximal}  \hspace{0.5cm} 
& \leftarrow \mbox{Seesaw enhacement}  \hspace{0.5cm} & 
\hspace{0.5cm} V_{up} = I \\
U_{lepton} = V_{CKM}
\hspace{0.5cm} 
& \leftarrow \mbox{GUT} \rightarrow \hspace{0.5cm} &
V_{down} = V_{CKM}
\end{eqnarray}
where ``Seesaw enhacement'' indicates that the mechanism my be 
responsible for neutrino-origin bimaximal (or bi-large) matrix 
\cite{seesaw_enhance}. 
Notice that while the maximal mixing comes purely from the lepton 
sector in these constructions, an amalgam of quark and lepton mixing 
arises once the GUT constraint is imposed.

Now we briefly review these scenarios and their consequences in a minimal way; 
See \cite{mina-smi} for more detailed discussions. 
In the lepton-origin bimaximal scenario, the MNS matrix can be written as 
\begin{eqnarray}
U_{MNS}  = R_{23}^m \Gamma_{\delta} R_{12}^m  V^{CKM \dagger} = 
R_{23}^m \Gamma_{\beta} 
R_{12} (\pi/4 - \theta_{12}^{CKM}) 
R_{13}^{CKM \dagger} 
R_{23}^{CKM \dagger}. 
\label{lepton-origin}
\end{eqnarray}
where $\Gamma_{\delta} = \mbox{diag} [1, 1, e^{i \delta}]$.  
The lepton-origin bimaximal scenario is also discussed in \cite{raidal}.
Whereas in the neutrino-origin bimaximal scenario, 
it takes the form
\begin{eqnarray}
U_{MNS}  = V^{CKM \dagger}\Gamma_\delta  R_{23}^m R_{12}^m =
R_{12}^{CKM \dagger} R_{13}^{CKM \dagger} R_{23}^{CKM \dagger} 
\Gamma_{\delta} 
R_{23}^m R_{12}^m , 
\label{neutrino-origin}
\end{eqnarray}
It is worth to note that the order of rotations and the location 
where the maximal angle is inserted deserve careful attention 
\cite{mina-smi}.


\begin{table}
\vglue 0.5cm
\small
\begin{tabular}{|c|c|c|c|c|}
\hline
       & \hspace{0.1cm} $\Delta \sin^2 \theta_{12}$ \hspace{0.1cm} & \hspace{0.1cm} $\sin^2 2\theta_{13}$ \hspace{0.1cm} &  $D_{23} \equiv \frac{1}{2}-s^2_{23}$ \hspace{0.0cm} & \hspace{0.1cm} $J_{lep}/\sin \delta$ \hspace{0.1cm} \\
Scenarios      &             &                      &        &      \\
\hline
neutrino bi-maximal & 0.051 & 0.10 $\pm$ 0.032 & 0.025  & $1.5\times 
10^{-3}$  \\
lepton bi-maximal & $-6\times 10^{-4}$ & $2\times 10^{-3}$ & 0.035$^*$ & 
$5\times 10^{-3}$  \\
hybrid bi-maximal & $1.4\times 10^{-4}$ & $3.3\times 10^{-4}$ & 0.04$^*$ & 
$2.1\times 10^{-3}$     \\
neutrino max+large & 0.057 $\pm$ 0.023  & 0.10 $\pm$ 0.032 & SK bound & $\leq 6.8\times 10^{-3}$  \\
lepton max+large & $-6\times 10^{-4}$ & $2\times 10^{-3}$ & SK bound & $\leq 5\times 10^{-3}$   \\
hybrid max+large & $1.4\times 10^{-4}$ & $3.3\times 10^{-4}$ & SK bound & $\leq 2.1\times 10^{-3}$ \\
single maximal & 0.015  & 0.034 & $0.06 - 0.16$  &  $9.1\times 10^{-3}$  \\
\hline
\end{tabular}
\vglue 0.5cm

\caption[aaa]{
Predictions to the deviation from the QLC relation 
$\Delta \sin^2 \theta_{12}$, $\sin^2 2\theta_{13}$, the deviation 
parameter from the maximal 2-3 mixing $D_{23}$, and 
the leptonic Jarlskog factor $J_{lep}$ for different scenarios. 
The uncertainties indicated with $\pm$ 
come from the experimental uncertainty of the  atmospheric 
mixing angle $\theta_{23}$. 
Whenever there exist uncertainty due to the CP violating phase 
$\delta$ we assume that $\cos \delta = 0$ to obtain an 
``average value''.  For the quantities 
 which vanish at $\cos \delta = 0$ (indicated by *) 
the numbers are calculated by assuming 
$\cos \delta = 1$ 
``SK bound''  implies 
the whole region allowed by the Super-Kamiokande:  
$|D_{23}| \leq 0.16$. 
The numbers for the last row (single-maximal case) are computed 
with the assumed values of $\theta^l_{23}=\theta_{C}$ and 
$\theta^{\nu}_{23}=27^{\circ}$. 
}

\vglue 0.2cm
\label{table1}
\end{table}

Having specified the MNS matrix it is straightforward to work out the 
phenomenological consequences. 
Instead of repeating the discussion given in  \cite{mina-smi}, 
we give a summary Table~\ref{table1}. 
We define the parameter which describes deviation from 
the QLC relation (\ref{QLC}) as 
\begin{eqnarray}
\Delta  \sin^2\theta_{12} \equiv 
\sin^2\theta_{\odot} -
\sin^2 \left( \frac{\pi}{4} - \theta_C \right). 
\label{compare}
\end{eqnarray}
At the moment, $\Delta  \sin^2\theta_{12} = 0.002 \pm 0.040$ 
experimentally. 
Let us focus on the neutrino- and the lepton-origin bimaximal scenarios, 
the first and the second rows in Table~\ref{table1}. 
It should be noticed that there are the characteristic differences between them; 
In the lepton-origin bimaximal scenario, the deviation from 
the QLC relation (\ref{QLC}) is extremely small so that it is very 
difficult, if not impossible, to verify it experimentally. 
In the neutrino-origin bimaximal scenario, on the other hand, 
the deviation is sizable and may be in reach in the future solar 
and the reactor neutrino experiments. 
We will discuss in Sec.~7 how the accuracy of testing 
the QLC relation can be improved.

The readers might be surprised by proliferation of scenarios in 
Table~\ref{table1}. In addition to the first and the second rows that 
are discussed above there exist five more scenarios. 
It is because the QLC relation (\ref{QLC}) is satisfied at least approximately 
by scenarios with a single maximal angle in 1-2 sector. 
Therefore, there exist much wider possibilities, as given in Table~\ref{table1}. 
I stop here, leaving examination of these scenarios \cite{mina-smi} 
for interested readers.

\section{Renormalization Stability}

Suppose that there exists a GUT model which embodies the QLC 
relation. It is not quite sufficient to guarantee the QLC 
relation (\ref{solar-cabibbo}) at low energies,  
because the renormalization flow could destroy 
the relationship. It is known that the running of the Cabibbo angle 
is negligibly small in the SM and in the MSSM. 
For instance, in MSSM with $\tan\beta =  50$
the  parameter  $\sin \theta_C$ decreases from 0.2225 at the 
$m_Z$ down to 0.2224 at the $10^{16}$  GeV~\cite{C-running}.

The issue is, therefore, located in the running of leptonic angle $\theta_{12}$. 
The renormalization effect on the leptonic $\theta_{12}$ has been 
investigated by many people~\cite{renorm}.
It depends on the type of mass spectrum of light neutrinos. 
For the  spectrum with normal mass hierarchy, 
$m_1 < m_2 \ll m_3$, the effect is negligible. 
In contrast, in the case of quasi-degenerate spectrum, 
$m_1 \approx  m_2 \approx m_3 = m_0$,  or the spectrum with
inverted mass hierarchy the effects can be large. 
The most recent analysis of the renormalization effects in mixing parameters 
\cite{lindner} reassures that in most of the parameter space the QLC 
relation (\ref{QLC}) is stable under the renormalization flow.

\section{Quark-Lepton Mass Models with QLC Relation}

It is important to construct concrete models to which the QLC 
relation (\ref{QLC}) is embedded in a natural way. 
Let me describe a possible idea toward this direction by abstracting 
an essence from the detailed discussion given in \cite{mohapatra1}. 
Suppose that one can prepare a zeroth-order model in which the 
lepton and quark mixing matrix have the following form  
\begin{eqnarray}
U_{MNS} = U_{bimax}, 
\hspace{1cm}
V_{CKM} = 1.
\label{0th}
\end{eqnarray}
Then, one envisage the mechanism that generates the first-order 
correction to the leading-order formula such that it modifies 
(\ref{0th}) by the same amount given by the Cabibbo rotation, 
\begin{eqnarray}
U_{MNS} = U_{bimax} \times V_{Cabibbo-like}, 
\hspace{1cm}
V_{CKM} = 1 \times V_{Cabibbo-like},
\label{1st}
\end{eqnarray}
where $V_{Cabibbo-like}$ denotes the rotation only in 1-2 subspace 
by the amount of $\simeq \theta_{C}$ the Cabibbo angle. 
We note that it belongs to the neutrino-origin bimaximal scenario 
in the classification above.

Of course, the real question is if one can construct such a model 
as that it possesses the desirable zeroth-order structure and is 
able to generate the first-order corrections of the required form. 
The authors of \cite{mohapatra1} presented a model based on 
the Pati-Salam gauge group 
SU(2)$_{\mbox{L}} \times$ 
SU(2)$_{\mbox{R}} \times$ 
SU(4)$_{\mbox{c}}$,  
and presented arguments that it satisfies the above requirements. 
Since they are quite involved, I urge the interested readers to 
go to their paper \cite{mohapatra1}.

We note, in passing, that once the MNS matrix is written in the form 
as in (\ref{1st}) it is identical to the parametrization of lepton mixing 
matrix which is examined by many authors in much more generic 
context than the QLC relation \cite{ramond,parametrization}.

One may ask: 
``To which point we have reached and where to go?'', which may 
be too premature question to ask. 
I feel at this point that we still lack simple models in which the QLC relation 
is naturally implemented. Or, there might be a mechanism that can be 
called as ``built-in stability'' which remains to be understood. 
After the Neutrino Telescope workshop several papers related to 
QLC were submitted on the Archiv. 
The authors of \cite{mohapatra2} attempt a systematic search for 
higher dimensional operators which lead to the QLC relation 
within the framework of inverted mass hierarchy. 
Whereas in \cite{LSS} a mechanism called ``screening'' 
is proposed to prevent Dirac flavor structure from contaminating 
to the lepton mixing.

\section{Experimental Test of the QLC Relation}
\label{test}

We now discuss how the QLC relation (\ref{QLC}) can be tested 
experimentally. Since the Cabibbo angle is measured in an enormous 
precision as emphasized earlier, the real problem is to what accuracy the 
solar angle $\theta_{12}$ can be measured experimentally.
At this moment there exist two approaches to measure $\theta_{12}$ accurately. 
The first one is a natural extension of  the method by which 
$\theta_{12}$ is determined with the highest precision today, namely, 
combining the solar and the KamLAND experiments. 
The other one is to create a dedicated new reactor experiment with 
detector at around the first oscillation maximum of reactor neutrino oscillation. 
Let me briefly explain about the basic ideas behind them one by one.

\subsection{Solar-KamLAND method}

Combining the solar and the KamLAND experiments is powerful 
because solar neutrino measurement is good at constraining 
$\theta_{12}$ and KamLAND determines with high precision 
the other parameter $\Delta m^2_{21}$,  which makes 
the solar neutrino analysis essentially 1-dimensional. 
The former characteristics is particularly clear from the fact that 
the ratio of CC to NC rates in SNO directly determines 
$\sin^2{\theta}_{12}$ in the LMA solution. 
The current data allows accuracy of determination of $\sin^2{\theta}_{12}$ 
of about $\sim 15$\% (2 DOF) \cite{KamLAND_new}.
Further progress in measurement in SNO and KamLAND may 
improve the accuracy by a factor of $\sim$ 2 but not too much beyond that.

\begin{figure}[h]
\begin{center}
\vspace{-2.4cm}
\epsfig{figure=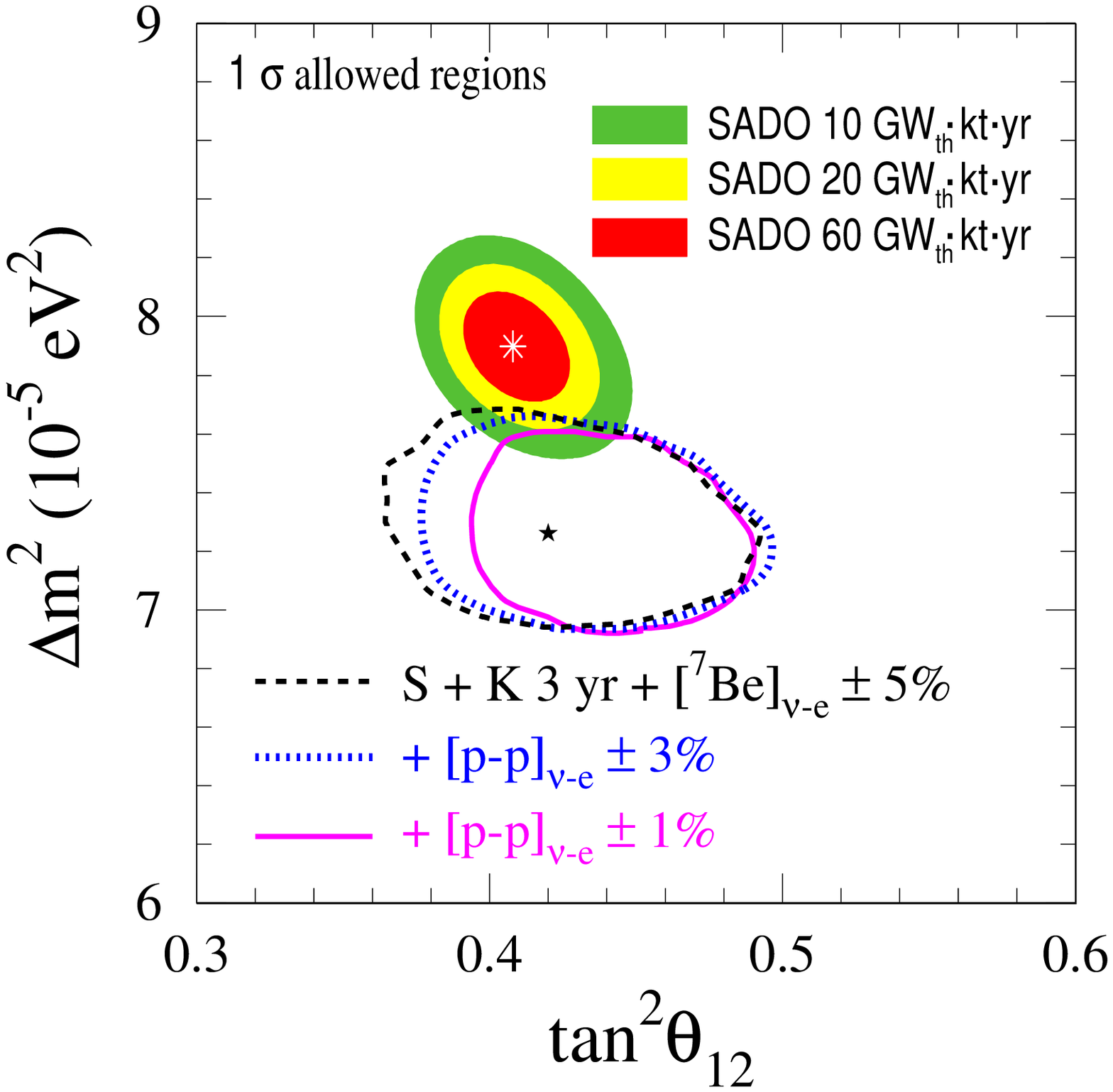,width=12cm}
\end{center}
\vspace{-4cm}
\caption{SADO's sensitivity contours are plotted in 
$\tan^2\theta_{12}$-$\Delta m^2_{21}$ space and are 
overlaid on Fig.6 of the roadmap paper by 
Bahcall and C.~Pe{\~n}a-Garay, 
in which the sensitivities 
of solar-KamLAND combined method are presented. 
The errors are defined both with 2 DOF.
}
\vspace{-1cm}
\label{SADOvsSK}
\end{figure}

However, if one want to improve substantially the accuracy of 
$\theta_{12}$ determination, the existing solar neutrino experiments 
are not quite enough. 
Measurement of low-energy $^7$Be and the pp neutrinos is 
particularly useful by exploring vacuum oscillation regime. 
The improvement made possible by these additional measurement is 
thoroughly discussed by Bahcall and Pe{\~n}a-Garay \cite{bahcall-pena}. 
Since the vacuum oscillation is the dominant mechanism at low energies 
measuring pp neutrino rate gives nothing but 
measurement of $\sin^2{2\theta}_{12}$. 
On the other hand, $^7$Be neutrino may carry unique informations 
of oscillation parameters due to its characteristic feature of 
monochromatic energy. 
The solar-KamLAND method will allow us to determine $\sin^2{\theta}_{12}$ 
to 4\% level at 1$\sigma$ CL \cite{bahcall-pena}. 
In the upper panels of Table~\ref{sensitivity}, 
we tabulate the sensitivities (1 DOF) 
currently obtained and expected by the future measurement. 
We show in Fig.~\ref{SADOvsSK} the contour of sensitivity expected by 
the method in the two-dimensional space spanned by 
$\tan^2{\theta}_{12}$ and $\Delta m^2_{21}$.

Fortunately, varying proposal  for such low energy solar neutrino 
measurement are available in the world \cite{nakahata}. 
Measurement of $^7$Be neutrinos is attempted in Borexino \cite{borexino} 
and in KamLAND \cite{inoue}.


\begin{table}
\vglue 0.2cm
\small
\begin{tabular}{c|cc}
\hline
\ Experiments\  & \ $\delta s^2_{12}/s^2_{12}$  at 68.27\% CL \ 
                & \ $\delta s^2_{12}/s^2_{12}$  at 99.73\% CL \  \\
\hline
 Solar+ KL (present)  & $ 8 $ \%  
                      & $ 26  $ \%   \\
\hline
 Solar+ KL (3 yr)  & $ 7 $ \% 
                   & $ 20 $ \%   \\
\hline
 Solar+ KL (3 yr) + pp (1\%) &  $ 4 $ \%  
                 & $ 11$ \%  \\
\hline
\multicolumn{3}{c}{54 km}\\
\hline
 SADO  for 10 $\mbox{GWth} \cdot$kt$\cdot$yr  &  4.6 \% \   (5.0 \%)  
                 &  12.2 \%  \ (12.9 \%) \\
\hline
 SADO for 20 $\mbox{GWth}  \cdot$kt$\cdot$yr
                 &  3.4 \% \  (3.8 \%)
                 &  8.8 \% \  (9.5 \%) \\
\hline
 SADO for 60 $\mbox{GWth}  \cdot$kt$\cdot$yr
                 & 2.1 \%  \   (2.4 \%)
                 & 5.5  \% \   (6.2 \%) \\
\hline
\end{tabular}
\vglue 0.4cm
\caption[aaa]{
Comparisons of fractional errors of the experimentally determined mixing angle, 
$\delta s^2_{12}/s^2_{12} \equiv \delta (\sin^2\theta_{12}) / \sin^2\theta_{12}$, 
by current and future solar neutrino experiments and KamLAND (KL), 
obtained from Tables 3 and 8 of Ref.~\cite{bahcall-pena}, versus that by SADO$_{\mbox{single}}$, which means to ignore all the other reactors 
than Kashiwazaki-Kariwa, obtained at 68.27\% 
and 99.73\% CL for 1 DOF in \cite{MNTZ}.
The numbers in parentheses are for SADO$_{\mbox{multi}}$, which takes 
into account all 16 reactors all over Japan.
}
\vglue 0.2cm
\label{sensitivity}
\end{table}

\subsection{SADO; Several-tens of km Antineutrino DetectOr}

Though natural and profitable as a dual-purpose experiment for 
both $\theta_{12}$ and solar flux measurement the solar-KamLAND 
method is not the unique possibility  for reaching the region of the 
highest sensitivity for $\theta_{12}$. 
The most traditional way of measuring mixing angles at the highest 
possible sensitivities is either to tune beam energy to the oscillation 
maximum (for example \cite{JPARC} which is for $\sin^2{2\theta}_{23}$), 
or to set up a detector at baseline corresponding to it as employed 
by various reactor experiments to measure $\theta_{13}$ \cite{reactor}. 
It is also notable that the first proposal of prototype superbeam 
experiment for detecting CP violation \cite{lowECP} entailed 
in a setup at around the first oscillation maximum.

For $\theta_{12}$ the latter method should apply to reactor neutrinos 
and in fact a concrete idea for a experimental proposal of dedicated 
reactor $\theta_{12}$ is worked out in detail \cite{MNTZ}.
See also \cite{goswami} for a similar but different proposal. 
The type of experiment is dubbed as  ``SADO'', an acronym of 
{\it Several-tens of km Antineutrino DetectOr} because of the range of 
baseline distance appropriate for the experiments \cite{MNTZ}.
It is a very feasible experiment because it does not require 
extreme reduction of the systematic error to 1\% level, 
as required in the $\theta_{13}$ measurement mentioned above. 
As is demonstrated in \cite{MNTZ} reduction of the systematic error to 
4\% level would be sufficient if no energy spectrum cut at 
$E_{prompt} = 2.6$ MeV is performed. 
It should be within reach in view of the current KamLAND error  
of 6.5\% \cite{KamLAND_new}. 
The effect of geo-neutrino background, which then has to be worried 
about without spectrum cut, is shown to be tolerable 
even for most conservative choice of geo-neutrino model, 
the Fully Radiogenic model  \cite{MNTZ}.

The accuracy achievable by the dedicated reactor $\theta_{12}$ 
measurement is quite remarkable. It will reach to  
2\% level at 1$\sigma$ CL for 60 
GW$_{\mbox{th}}$$\cdot \mbox{kt} \cdot \mbox{yr}$ exposure 
as shown in Table~\ref{sensitivity}. 
With Kashiwazaki-Kariwa nuclear reactor complex, it corresponds 
to about 6 years operation for KamLAND size detector. 
It is notable that possible uncertainty that may arise from the 
surrounding reactors besides the principle one is also modest,  
as one can see in Table~\ref{sensitivity}.

Notice that the measurement is not yet systematics dominated and 
therefore further improvement of the sensitivity is possible by 
gaining more statistics. 
In Fig.~\ref{SADOvsSK}, we make a comparison between 
the extended solar-KamLAND method and SADO single setup. 
If SADO can run long enough it can go beyond the solar-KamLAND method.

\section{Conclusion}

In this review talk, I introduced a new approach called 
 ``quark-lepton complementarity'' which has initiated on the impact of 
 the fascinating empirical relationship (\ref{solar-cabibbo}) 
 obtained as a result of numerous experiments supported by uncountable 
 numbers of people. 
I tried to sketch the ideas currently at hand which has been 
suggested in seeking deeper structure of quark-lepton relation 
through the the QLC relation. 
Most of the approach so far involves the maximal mixing in the 
1-2 sector. It is one of the most important step to understand 
the nature of the 1-2 maximal angle.

I have also discussed possible ways of testing the QLC relation 
(\ref{QLC}) to uncover deviation from it. As I have discussed 
it may testify from which sector, neutrinos or charged lepton, 
the 1-2 maximal angle originates. 
It is good to know that measurement of a few \% level accuracies 
in $\sin^2{\theta_{12}}$ is certainly possible either by an extended 
solar-KamLAND method, or the dedicated reactor $\theta_{12}$ 
experiment, SADO.

I have not covered in my talk possible role of the other mixing angles, 
$\theta_{23}$ and $\theta_{13}$. It is not understood if they are 
the registered members of the QLC  fraternity. 
Yet, since most approach to QLC anticipate nearly maximal 
$\theta_{23}$, possible deviation from its maximality would be 
of great interests. 
It should be noticed that 
the 1\% level measurement of $\sin^2{2\theta_{23}}$ does not 
translate into the similar sensitivity in $\sin^2{\theta_{23}}$. 
It is due to a large Jacobian involved in the transformation 
near the maximal angle, and the $\theta_{23}$ octant degeneracy. 
For detailed discussions of this point and for possible ways out, 
see \cite{MSS} and the references cited therein.

\section{Acknowledgements}

I thank Milla Baldo Ceolin for her cordial invitation to the by now 
traditional workshop series in Venice.
I was benefited from discussions with Morimitsu Tanimoto 
on various aspects of flavor physics. 
This work was supported in part by the Grant-in-Aid for Scientific Research, 
No. 16340078, Japan Society for the Promotion of Science.


\begin{thebibliography}{99}


\bibitem {SKatm_new}
Y.~Ashie {\it et al.}  [Super-Kamiokande Collaboration],
Phys.\ Rev.\ Lett.\  {\bf 93} (2004) 101801
[arXiv:hep-ex/0404034];


\bibitem{KamLAND_new}
T. ~Araki {\it et al.} [KamLAND Collaboration],
Phys.\ Rev.\ Lett.\  {\bf 94}, 081801 (2005) 
[arXiv:hep-ex/0406035].


\bibitem {K2K_new}
E.~Aliu {\it et al.}  [K2K Collaboration],
Phys.\ Rev.\ Lett.\  {\bf 94}, 081802 (2005) 
[arXiv:hep-ex/0411038].


\bibitem {SKatm}
Y.~Fukuda {\it et al.}  [Kamiokande Collaboration],
Phys.\ Lett.\ B {\bf 335}, 237 (1994);
%
Y.~Fukuda {\it et al.}  [Super-Kamiokande Collaboration],
Phys.\ Rev.\ Lett.\  {\bf 81}, 1562 (1998)
[arXiv:hep-ex/9807003];
Y.~Ashie {\it et al.}  [Super-Kamiokande Collaboration],
  arXiv:hep-ex/0501064.


\bibitem {MNS}
Z.~Maki, M.~Nakagawa and S.~Sakata,
Prog.\ Theor.\ Phys.\  {\bf 28}, 870 (1962).
See also, B.~Pontecorvo, 
Zh. Eksp. Teor. Fyz. {\bf 53}, 1717 (1967) 
[Sov. Phys. JETP {\bf 26}, 984 (1968)].  


\bibitem {solar}
B.~T.~Cleveland {\it et al.},
Astrophys.\ J.\  {\bf 496}, 505 (1998);
%
J.~N.~Abdurashitov {\it et al.}  [SAGE Collaboration],
Phys.\ Rev.\ C {\bf 60}, 055801 (1999)
[arXiv:astro-ph/9907113];
%
W.~Hampel {\it et al.}  [GALLEX Collaboration],
Phys.\ Lett.\ B {\bf 447}, 127 (1999);
%
S.~Fukuda {\it et al.}  [Super-Kamiokande Collaboration],
  Phys.\ Lett.\ B {\bf 539}, 179 (2002)
  [arXiv:hep-ex/0205075];
%
M.~B.~Smy {\it et al.}  [Super-Kamiokande Collaboration],
  Phys.\ Rev.\ D {\bf 69}, 011104 (2004)
  [arXiv:hep-ex/0309011];
%
Q.~R.~Ahmad {\it et al.}  [SNO Collaboration],
Phys.\ Rev.\ Lett.\  {\bf 87}, 071301 (2001)
[arXiv:nucl-ex/0106015];
{\it ibid.} {\bf 89}, 011301 (2002)
[arXiv:nucl-ex/0204008]; 
B.~Aharmim {\it et al.}  [SNO Collaboration],
  arXiv:nucl-ex/0502021.


\bibitem {CHOOZ}
M.~Apollonio {\it et al.}  [CHOOZ Collaboration],
  Phys.\ Lett.\ B {\bf 466}, 415 (1999)
  [arXiv:hep-ex/9907037].
See also, The Palo Verde Collaboration,
F.~Boehm {\it et al.},
Phys.\ Rev.\ D {\bf 64} (2001) 112001 
[arXiv:hep-ex/0107009].


\bibitem{QLC_old}
A.~Gamba, R.~E.~Marshak and S.~Okubo, 
Proc. Nat. Acad. Sci. {\bf 45}, 881 (1959); 
Z.~Maki, M.~Nakagawa, Y.~Ohnuki, and S.~Sakata,
Prog. Theor. Phys.  {\bf 23}, 1174 (1960).


\bibitem{GUT}
J.~C. Pati and A.~Salam, Phys. Rev. \textbf{D10} (1974), 275.
%
H.~Georgi and S.~L. Glashow, Phys. Rev. Lett. \textbf{32} (1974), 438.


\bibitem{mina-smi} 
H.~Minakata and A.~Yu Smirnov, 
Phys.\ Rev.\ D {\bf 70}, 073009 (2004)
[arXiv:hep-ph/0405088].


\bibitem{raidal} 
M. Raidal, Phys.\ Rev.\ Lett.\  {\bf 93}, 161801 (2004)
[arXiv:hep-ph/0404046]. 


\bibitem{mohapatra} 
The other aspects of the quark-lepton relation may be found e.g., in a talk 
presented by Mohapatra; 
 R.~N.~Mohapatra,
  arXiv:hep-ph/0504138.



\bibitem{NOVE03_smi}
The first public statement of this relation, to me knowledge, 
was done by Alexei Smirnov in his talk given at 2nd International Workshop 
on Neutrino Oscillations in Venice (NO-VE 2003), Venice, Italy, 3-5 Dec 2003. 
  A.~Y.~Smirnov,
  arXiv:hep-ph/0402264.


\bibitem{pakvasa}
J.~Ferrandis and S.~Pakvasa,
Phys.\ Rev.\ D {\bf 71}, 033004 (2005)
[arXiv:hep-ph/0412038].


\bibitem{mohapatra1}
P.~H.~Frampton and R.~N.~Mohapatra,
JHEP {\bf 0501}, 025 (2005)
[arXiv:hep-ph/0407139].


\bibitem{ramond}
A.~Datta, L.~Everett and P.~Ramond,
arXiv:hep-ph/0503222.


\bibitem{QLCmore}
  S.~K.~Kang, C.~S.~Kim and J.~Lee,
  arXiv:hep-ph/0501029; 
  N.~Li and B.~Q.~Ma,
  arXiv:hep-ph/0501226;
  K.~Cheung, S.~K.~Kang, C.~S.~Kim and J.~Lee,
  arXiv:hep-ph/0503122;
  Z.~z.~Xing,
  arXiv:hep-ph/0503200; 


\bibitem{mohapatra2}
S.~Antusch, S.~F.~King and R.~N.~Mohapatra,
arXiv:hep-ph/0504007.


\bibitem{LSS}
M.~Lindner, M.~A.~Schmidt and A.~Y.~Smirnov,
arXiv:hep-ph/0505067.


\bibitem{PDG}
S.~Eidelman {\it et al.}  [Particle Data Group Collaboration],
Phys.\ Lett.\ B {\bf 592} 1 (2004) 1. 


\bibitem{global}
M.~Maltoni, T.~Schwetz, M.~A.~Tortola, and J.~W.~F.Valle,
New J. Phys.  {\bf 6}  (2004) 122.
[arXiv:hep-ph/0405172]. 


\bibitem{mu-tau}
K.~S.~Babu, E.~Ma, and J.~W.~F.~Valle, 
Phys.\ Lett.\ B {\bf 552}, 207 (2003) 
[hep-ph/0206292];
%
 E.~Ma,
  Phys.\ Rev.\ D {\bf 66}, 117301 (2002)
  [arXiv:hep-ph/0207352];
Mod.\ Phys.\ Lett.\ A {\bf 17}, 2361 (2003) 
[hep-ph/0211393];
E.~Ma and G.~Rajasekaran,
  Phys.\ Rev.\ D {\bf 68}, 071302 (2003)
  [arXiv:hep-ph/0306264]; 
%
W.~Grimus and L.~Lavoura, 
Phys.\ Lett.\ B {\bf 572}, 189 (2003)
[arXiv:hep-ph/0305046];
%
Acta.\ Phys.\ Polon.\ {\bf B34}, 5393 (2003) 
[arXiv:hep-ph/0310050];
%
J.~Kubo, A.~Mondragon, M.~Mondragon and E.~Rodriguez-Jauregui,
Prog.\ Theor.\ Phys.\  {\bf 109}, 795 (2003)
[arXiv:hep-ph/0302196]; 
J.~Kubo,
Phys.\ Lett.\ B {\bf 578}, 156 (2004)
[arXiv:hep-ph/0309167];
%
R.~N.~Mohapatra,
  JHEP {\bf 0410}, 027 (2004)
  [arXiv:hep-ph/0408187].





\bibitem{bimax}
F.~Vissani,  (1997),  hep-ph/9708483;
%
V.~D. Barger, S.~Pakvasa, T.~J. Weiler, and K.~Whisnant, Phys. Lett.
  \textbf{B437} (1998), 107,  [hep-ph/9806387];
%
A.~J. Baltz, A.~S. Goldhaber, and M.~Goldhaber, Phys. Rev. Lett. \textbf{81}
  (1998), 5730,  [hep-ph/9806540];
%
H.~Georgi and S.~L. Glashow, Phys. Rev. \textbf{D61} (2000), 097301,
  [hep-ph/9808293];
%
I.~Stancu and D.~V. Ahluwalia, Phys. Lett. \textbf{B460} (1999), 431,
  [hep-ph/9903408].


\bibitem{e-mu-tau}
S.~T.~Petcov,
  Phys.\ Lett.\ B {\bf 110} (1982) 245; 
R.~Barbieri, L.~J.~Hall, D.~R.~Smith, A.~Strumia and N.~Weiner,
  JHEP {\bf 9812}, 017 (1998)
  [arXiv:hep-ph/9807235]; 
A.~S.~Joshipura and S.~D.~Rindani,
  Eur.\ Phys.\ J.\ C {\bf 14}, 85 (2000)
  [arXiv:hep-ph/9811252];
R.~N.~Mohapatra, A.~Perez-Lorenzana and C.~A.~de Sousa Pires,
  Phys.\ Lett.\ B {\bf 474}, 355 (2000)
  [arXiv:hep-ph/9911395]; 
 T.~Kitabayashi and M.~Yasue,
  Phys.\ Rev.\ D {\bf 63}, 095002 (2001)
  [arXiv:hep-ph/0010087]; 
  Phys.\ Lett.\ B {\bf 524}, 308 (2002)
  [arXiv:hep-ph/0110303];
W.~Grimus and L.~Lavoura,
  Phys.\ Rev.\ D {\bf 62}, 093012 (2000)
  [arXiv:hep-ph/0007011];
  JHEP {\bf 0107}, 045 (2001)
  [arXiv:hep-ph/0105212]; 
  R.~N.~Mohapatra,
  Phys.\ Rev.\ D {\bf 64}, 091301 (2001)
  [arXiv:hep-ph/0107264];
K.~S.~Babu and R.~N.~Mohapatra,
  Phys.\ Lett.\ B {\bf 532}, 77 (2002)
  [arXiv:hep-ph/0201176];
 H.~S.~Goh, R.~N.~Mohapatra and S.~P.~Ng,
  Phys.\ Lett.\ B {\bf 542}, 116 (2002)
  [arXiv:hep-ph/0205131];
 H.~J.~He, D.~A.~Dicus and J.~N.~Ng,
  Phys.\ Lett.\ B {\bf 536}, 83 (2002)
  [arXiv:hep-ph/0203237];
Q.~Shafi and Z.~Tavartkiladze,
  Phys.\ Lett.\ B {\bf 482}, 145 (2000)
  [arXiv:hep-ph/0002150].


\bibitem{lopsided}
K.~S. Babu and S.~M. Barr, Phys. Lett. {\bf  B381} (1996), 202,
  [arXiv:hep-ph/9511446];
%
S.~M. Barr, Phys. Rev. {\bf  D55} (1997), 1659,  [arXiv:hep-ph/9607419];
%
J.~Sato and T.~Yanagida, 
Phys. Lett. B {\bf 430}, 127 (1998) [arXiv:hep-ph/9710516]; 
Nucl. Phys. Proc. Suppl. {\bf 77}, 293 (1999) [arXiv:hep-ph/9809307];
C.~H. Albright and S.~M. Barr, Phys. Rev. {\bf  D58} (1998), 013002,
  [arXiv:hep-ph/9712488];
%
G.~Altarelli and F.~Feruglio, Phys. Lett. {\bf B439} (1998), 112,
  [arXiv:hep-ph/9807353];
%
G.~Altarelli and F.~Feruglio, JHEP {\bf 11} (1998), 021,  [arXiv:hep-ph/9809596].


\bibitem{seesaw_enhance}
A. Yu. Smirnov, Phys. Rev. {\bf D48}, 3264 (1993).
[arXiv:hep-ph/9304205]; 
%
M.~Tanimoto, 
Phys. Lett. B {\bf 345}, 477 (1995) 
[arXiv:hep-ph/9503318];
%
G. Altarelli, F. Feruglio, and I. Masina, 
Phys. Lett. B {\bf 472}, 382 (2000) 
[arXiv: hep-ph/9907532];
T.~K. Kuo, G.-H. Wu, and S.~W. Mansour, Phys. Rev. {\bf D61} (2000), 111301,
   [arXiv:hep-ph/9912366];
%
G.~Altarelli, F.~Feruglio, and I.~Masina, Phys. Lett. {\bf B472} (2000),
  382,  [arXiv:hep-ph/9907532];
%
S.~Lavignac, I.~Masina, and C.~A. Savoy, Nucl. Phys. {\bf B633} (2002),
  139,  [arXiv:hep-ph/0202086];
%
A.~Datta, F.-S. Ling, and P.~Ramond, Nucl. Phys. {\bf B671} (2003),
  383,  [arXiv:hep-ph/0306002];
%
M.~Bando, S.~Kaneko, M.~Obara, and M.~Tanimoto, Phys. Lett. {\bf B580}
  (2004), 229,  [arXiv:hep-ph/0309310].









\bibitem{C-running}
H. Arason {\it et. al.,} Phys. Rev. D{\bf 46}, 3945 (1992).  



\bibitem{renorm}
J.~A.~Casas, J.~R.~Espinosa, A.~Ibarra, and I.~Navarro, 
Nucl. Phys. B \textbf{573}, 652 (2000) 
[hep-ph/9910420];
%
K.~R.~S.~Balaji, A.~S.~Dighe, R.~N.~Mohapatra, and M.~K.~Parida,
Phys. Lett. B \textbf{481}, 33 (2000) 
[hep-ph/0002177];
%
N.~Haba, Y.~Matsui, and N.~Okamura, 
Eur. Phys. J. C \textbf{17}, 513 (2000) 
[hep-ph/0005075];
%
P.~H.~Chankowski and S.~Pokorski, 
Int. J. Mod. Phys. A \textbf{17}, 575 (2002) 
[hep-ph/0110249];
S.~Antusch, J.~Kersten, M.~Lindner, and M.~Ratz, 
Nucl. Phys. B \textbf{674}, 401 (2003) 
[arXiv:hep-ph/0305273].


\bibitem{lindner}
S.~Antusch, J.~Kersten, M.~Lindner, M.~Ratz and M.~A.~Schmidt,
JHEP {\bf 0503}, 024 (2005)
[arXiv:hep-ph/0501272].




\bibitem{parametrization}
C. Giunti and M. Tanimoto, 
Phys.\ Rev.\ D {\bf 66}, 053013,(2002), 
{\it ibid} {\bf 66}, 113006 (2002) 
[hep-ph/0207096, hep-ph/0209169].
%
P. H. Frampton, S. T. Petcov, and W. Rodejohann,  
Nucl. Phys. B {\bf 687}, 31 (2004).  
[arXiv:hep-ph/0401206];
%
 G.~Altarelli, F.~Feruglio and I.~Masina,
  Nucl.\ Phys.\ B {\bf 689}, 157 (2004)
  [arXiv:hep-ph/0402155]; 
%
W.~Rodejohann,
  Phys.\ Rev.\ D {\bf 70}, 073010 (2004)
  [arXiv:hep-ph/0403236].




\bibitem{bahcall-pena} 
J.~N.~Bahcall and C.~Pe{\~n}a-Garay, 
JHEP {\bf 0311} (2003) 004.


\bibitem{nakahata}
M.~Nakahata, 
Talk at the 5th Workshop on 
``Neutrino Oscillations and their Origin'' (NOON2004), 
February 11-15, 2004, Odaiba, Tokyo, Japan.
http://www-sk.icrr.u-tokyo.ac.jp/noon2004/


\bibitem{borexino}
G.~Alimonti {\it et al.}  [Borexino Collaboration],
Astropart.\ Phys.\  {\bf 16} (2002) 205.


\bibitem{inoue}
K.~Inoue, Talk at Neutrino Oscillation Workshop (NOW2004), 
September 11-17, 2004, Conca Specchiulla, Otranto, Italy. 


\bibitem{JPARC}
Y.~Itow {\it et al.}, arXiv:hep-ex/0106019. 
An updated version: \\
http://neutrino.kek.jp/jhfnu/loi/loi.v2.030528.pdf 


\bibitem{reactor} 
See, for example, 
H.~Minakata, H.~Sugiyama, O.~Yasuda, K.~Inoue, and F.~Suekane, 
Phys.\ Rev.\ D {\bf 68}, 033017 (2003)
[arXiv:hep-ph/0211111]; 
K.~Anderson  {\it et al.},
White Paper Report on Using Nuclear Reactors to Search for a 
Value of $\theta_{13}$, arXiv:hep-ex/0402041. 


\bibitem{lowECP} 
H.~Minakata and H.~Nunokawa, Phys.\ Lett.\ {\bf B495} (2000) 369;
[arXiv:hep-ph/0004114];


\bibitem{MNTZ}
H.~Minakata, H.~Nunokawa, W.~J.~C.~Teves and R.~Zukanovich Funchal, 
Phys.\ Rev.\ D  {\bf 71}  (2004) 013005 
[arXiv:hep-ph/0407326];
arXiv:hep-ph/0501250.


\bibitem{goswami}
A.~Bandyopadhyay, S.~Choubey and S.~Goswami,
Phys.\ Rev.\ D {\bf 67}, 113011 (2003)
[arXiv:hep-ph/0302243]; 
A.~Bandyopadhyay, S.~Choubey, S.~Goswami and S.~T.~Petcov,
  arXiv:hep-ph/0410283.
  

\bibitem {MSS}
H.~Minakata, M.~Sonoyama and H.~Sugiyama, 
Phys.\ Rev.\ D {\bf 70},  113012 (2004) 
[arXiv:hep-ph/0406073]. 






\end{thebibliography}
\end{document}